\documentclass[aps, prd, twocolumn,superscriptaddress,nofootinbib]{revtex4-1}

\usepackage{latexsym,cancel,amssymb,amsmath,amsthm,verbatim,mathrsfs}
\usepackage{color,graphicx}

\usepackage[colorlinks,citecolor=blue]{hyperref}
\usepackage{slashed}

\newcommand{\beq}{\begin{equation}}
\newcommand{\eeq}{\end{equation}}
\newcommand{\bea}{\begin{eqnarray}}
\newcommand{\eea}{\end{eqnarray}}
\newcommand{\met}{\slashed{\rm E}_T}
\newcommand{\nn}{\nonumber}

\allowdisplaybreaks[4]
\begin{document}

\title{What can We Learn from Triple Top-Quark Production?}

\author{Qing-Hong Cao}
\email{qinghongcao@pku.edu.cn}
\affiliation{Department of Physics and State Key Laboratory 
of Nuclear Physics and Technology, Peking University, Beijing 100871, China}
\affiliation{Collaborative Innovation Center of Quantum Matter, Beijing 100871, China}
\affiliation{Center for High Energy Physics, Peking University, Beijing 100871, China}

\author{Shao-Long Chen}
\email{chensl@mail.ccnu.edu.cn}
\affiliation{Key Laboratory of Quark and Lepton Physics (MoE) and Institute of Particle Physics, Central China Normal University, Wuhan 430079, China}
\affiliation{Center for High Energy Physics, Peking University, Beijing 100871, China}

\author{Yandong Liu}
\email{ydliu@bnu.edu.cn}
\affiliation{Key Laboratory of Beam Technology of Ministry of Education, College of Nuclear Science and Technology, Beijing Normal University, Beijing 100875, China}
\affiliation{Beijing Radiation Center, Beijing 100875, China}

\author{Xiao-Ping Wang}
\email{xia.wang@anl.gov}
\affiliation{High Energy Physics Division, Argonne National Laboratory, Argonne, IL 60439, USA}

\begin{abstract}
Different from other multiple top-quark productions, triple top-quark production requires the presence of both flavor violating neutral interaction and flavor conserving neutral interaction. We describe the interaction of triple top-quarks and up-quark in terms of two dimension-6 operators; one can be induced by a new heavy vector resonance, the other by a scalar resonance. Combining same-sign top-quark pair production and four top-quark production, we explore the potential of the 13~TeV LHC on searching for the triple top-quark production. 
\end{abstract}

\maketitle

\noindent{\bf 1. Motivation.}

Searching for new physics (NP) beyond the Standard Model (SM) is the major task of the Large Hadron Collider (LHC). Many searching programs involve top-quark ($t$) which is commonly believed to be sensitive to NP at the TeV scale,  e.g.  opposite-sign or same-sign top-quark pair production, single top-quark production and four top-quark production. Unfortunately, triplet top-quark production is not paid too much attentions yet. We argue that the triple-top production is very unique among all the NP searching programs related to top-quarks as it is an undoubted signature of Flavor Violating Neutral Intearction (FVNI).

Although highly suppressed by Glashow-Iliopoulos-Maiani mechanism~\cite{Glashow:1970gm} in the SM, the FVNI effect can be sizable in many well motivated NP models; therefore, measuring the FVNI is commonly believed to be a good probe of NP beyond the SM. The FVNI effects in the lepton sector and light-quark mesons have been well tested and no clear evidents were reported yet. The FVNI of top-quark can only be tested in hadron collisions. For example, the top-quark FVNI vertices ($tq\gamma$, $tqg$ and $tqh$) have been probed in the top quark production or decay processes at the Tevatron and LHC \cite{Abe:1997fz,Aaltonen:2008ac,Aaltonen:2008qr,Abazov:2010qk,Abazov:2011qf,Aad:2012ij,Aad:2014dya,CMS:2014hwa,Chatrchyan:2013nwa}, and severe constraints on the NP generating such effects are obtained. It is in general difficult to directly probe the top-quark FVNI effect at the LHC as those FVNI couplings are extremely small.

Even worse, some searching programs of top FVNI effects can not fully confirm its existence.
For example, the same-sign top-quark pair ($tt/\bar{t}\bar{t}$) production~\cite{Berger:2011ua}, often thought as a gold channel of probing top quark FVNI interactions, can be mimicked by a color sextet scalar or vector~\cite{Berger:2010fy,Zhang:2010kr}.

\begin{figure}
\centering
\includegraphics[scale=0.35]{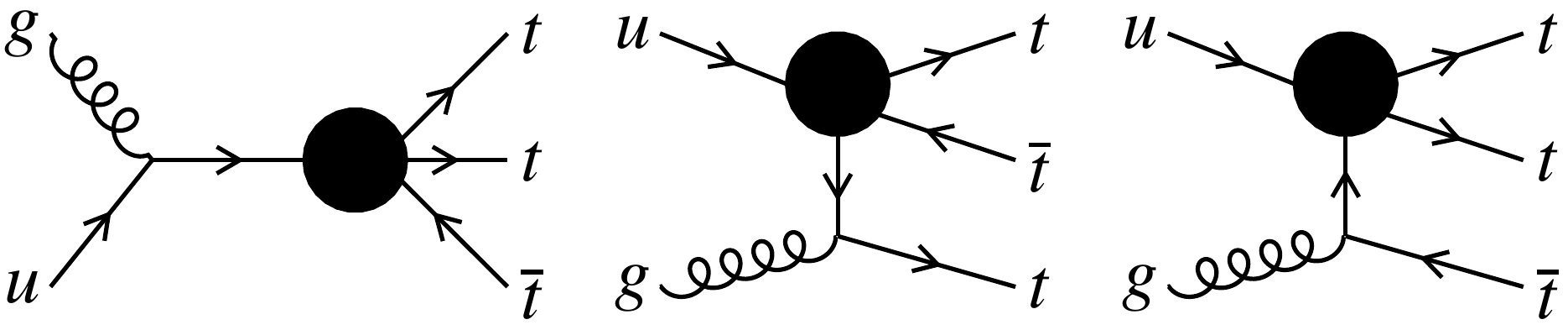}
\caption{\it Illustration of Feynman diagrams for the triple top quark production induced by the $uttt$ operator.}
\label{uttt_feyn}
\end{figure}

The triple top-quark production unambiguously points to the occurrence of top-quark FVNI. 
It can be understood from the charge conservation. 
As the top-quark has charge $+2/3$, the triple top-quarks in the final state have a electromagnetic charge of $\pm 2/3$ ($tt\bar{t}$ or $t\bar{t}\bar{t}$) or $\pm 2$ ($ttt$ or $\bar{t}\bar{t}\bar{t}$).
While the parton inside the initial proton has charge either $+2/3$ (up-type quark) or $-1/3$ (down-type quark), the maximal net charge in the initial state can be $\pm 4/3$. Therefore
the triple-top quark in the final state can be only in the form of $tt\bar{t}$ or $t\bar{t}\bar{t}$, 
which demands the initial state consists of  an up-type quark and a gluon. Due to absence of top quark as a parton inside the proton at the LHC, there must exist a FVNI interaction between the top-quark and up-type light quark in the triple top-quark production; see Fig.~\ref{uttt_feyn} for illustration. We consider only up-quark in this study as the charm-quark contribution is highly suppressed by the parton distribution functions. 

Another special feature of the triple top-quark production is that it also needs flavor conserving neutral interactions (FCNI). In a renormalizable theory only new heavy scalars or vectors can generate the triple top-quark production at the tree-level, and regardless of the detailed interaction form, the heavy scalar or vector must connect to the $t$-quark and $u$-quark on one side while to a pair of top-quarks on the other side.  The latter interaction conserves the top-quark flavor. Therefore, the triple top-quark production requires the FVNI and FCNI present simultaneously which make the channel very unique in top-quark physics.

In this work we explore the phenomenology of triple top-quark productions at the LHC and future 100~TeV collider. The triple top-quark production can be induced in many NP models which have extra heavy scalars or vector resonances with top-quark FVNI interactions~\cite{Barger:2010uw,Han:2012qu}. 
In this paper we will assume that such new scalar or vector effects are indeed present, but that the energies available at present and near-future colliders lie below their typical NP scale 􏰗$\Lambda$. In this case the characteristics of the new interactions can be probed only through their virtual effects on processes involving SM particles; such effects can be efficiently coded in a model-independent way using the well-studied effective-Lagrangian formalism~\cite{Weinberg:1978kz,Georgi:1991ch,Wudka:1994ny}. All new physics effects can be parameterized by the coefficients of a series of gauge-invariant operators ($\mathcal{O}_{i}$) constructed out of the SM fields; when the heavy physics decouples, these operators have dimensions $\ge 5$ and their coefficients are suppressed by inverse powers of the NP scale $\Lambda$.

In the study we consider the following two operators given in Ref.~\cite{Chen:2014ewl}: 
\begin{align} 
&\mathcal{O}^{\mathcal{S}}_{uttt} = \frac{c_\mathcal{S}}{\Lambda^2}(\bar{t}  t) (\bar{t}P_R u),  \label{s-type} \\
&\mathcal{O}^{\mathcal{V}}_{uttt} = \frac{c_\mathcal{V}}{\Lambda^2}(\bar{t}\gamma^\mu P_R t)(\bar{t}\gamma_\mu P_R u). \label{v-type}
\end{align}
The superscript of the operator denotes that the operator can be generated by a new heavy color-neutral scalar ($\mathcal{S}$) or vector ($\mathcal{V}$). Note that the operator $\mathcal{O}^{\mathcal{V}}_{uttt} $ can also be generated by a color-sextet scalar. For example, the $uttt$ operator induced by a color-sextet scalar reads as 
\begin{equation}
\frac{1}{\Lambda^2}(\delta^{ik}\delta^{jl}+\delta^{il}\delta^{jk})(\bar{t}^{c}_i P_R u_j) (\bar{t}_k P_L t^c_l),
\end{equation}
where $i,j,k,l$ are the color indexes of quarks. It yields $\mathcal{O}^{\mathcal{V}}_{uttt} $ after the Fierz transformation
\begin{align}
&(\bar{t}^c_iP_R u_j)(\bar{t}_k P_L t^c_l)   
=\frac{1}{2}(\bar{t}_l\gamma^\mu P_R t_i )(\bar{t}_k\gamma^\mu P_R u_j). 
\end{align}
We thus focus on the color-neutral operators throughout this study and the result can be easily extended to the color-sextet operator after a proper rescaling.

The scalar $\mathcal{S}$ or the vector $\mathcal{V}$  can also affect the same-sign top-quark pair ($tt/\bar{t}\bar{t}$) production and four top-quark ($t\bar{t}t\bar{t}$) production. We separate the FVNI and FCNI in the operator $\mathcal{O}_{uttt}^{\mathcal{S},\mathcal{V}}$ as follows: 
\begin{equation}
c_{\mathcal{S}} = f_{\rm FVNI}^{\mathcal{S}} f_{\rm FCNI}^{\mathcal{S}}, \qquad c_{\mathcal{V}} = f_{\rm FVNI}^{\mathcal{V}} f_{\rm FCNI}^{\mathcal{V}},
\end{equation}
where $f_{\rm V}^{\mathcal{S}(\mathcal{V})}$ and $f_{\rm C}^{\mathcal{S}(\mathcal{V})}$ describes the FVNI and FCNI induced by $\mathcal{S}$($\mathcal{V}$), respectively. 
The $tt/\bar{t}\bar{t}$ production can be affected by the FVNI through the following two operators:
\begin{align}
&\mathcal{O}^{\mathcal S}_{uutt} =\frac{1}{2} \frac{(f_{\rm FVNI}^{\mathcal{S}})^2}{\Lambda^2}(\bar{t}P_R u) (\bar{t}P_R u), \label{tts-type}  \\
&\mathcal{O}^{\mathcal V}_{uutt} = \frac{1}{2}\frac{(f_{\rm FVNI}^{\mathcal{V}})^2}{\Lambda^2}(\bar{t}\gamma^\mu P_R u)(\bar{t}\gamma_\mu P_R u). \label{ttv-type}
\end{align}
We assume that all the operators are induced all by a NP resonance at the tree-level such that they exhibit the same cutoff scale $\Lambda$. 
The FCNI naturally induces four top-quark effective operators as follows:
\begin{align}
&\mathcal{O}^{\mathcal{V}}_{tttt} = \frac{1}{2}\frac{(f_{\rm FCNI}^{\mathcal{V}})^2}{\Lambda^2}(\bar{t}\gamma^\mu P_R t)(\bar{t}\gamma_\mu P_R t), 
\label{v-type}
\\
&\mathcal{O}^{\mathcal{S}}_{tttt} = \frac{1}{2}\frac{(f_{\rm FCNI}^{\mathcal{S}})^2}{\Lambda^2}(\bar{t}   t) (\bar{t}  t).  \label{s-type} 
\end{align}
Such operators contribute to the $t\bar{t}t\bar{t}$ production at the LHC which can be utilized to measure the top quark Yukawa coupling directly \cite{Cao:2016wib}.

~\\
\noindent{\bf 2. Triple top-quark production.}

\begin{figure}[b]
\includegraphics[scale=0.45]{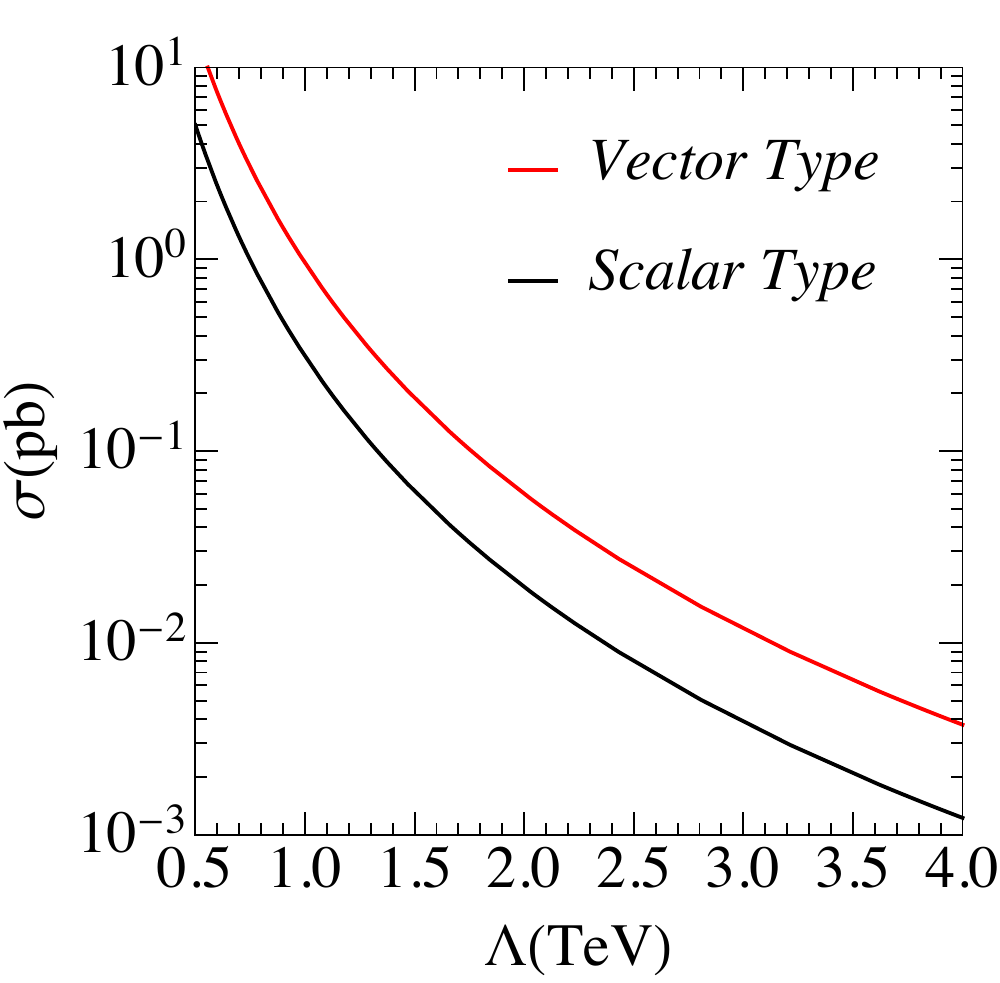}
\caption{\it Cross section of the triple top-quark production as a function of $\Lambda$ for $f_{\rm FVNI}^{\mathcal{V(S)}} f_{\rm FCNI}^{\mathcal{V(S)}}=1$ at the 13 TeV LHC . } 
\label{fig:xsec} 
\end{figure}

Next we first explore the potential of the LHC on searching for the triple top-quark production and then comment on the NP effects in the $tt/t\bar{t}$ and $t\bar{t}t\bar{t}$ production. 
Defining the cross section of the triple top-quark production as 
\begin{equation}
\sigma_{ttt}^{\mathcal{S}(\mathcal{V})}\equiv \sigma^{\mathcal{S}(\mathcal{V})}(tt\bar{t})+\sigma^{\mathcal{S}(\mathcal{V})}({t\bar{t}\bar{t}}),
\end{equation}
we obtain the leading order cross section at the 13~TeV LHC as follows:
\begin{align} 
\label{cs_ttt}
\sigma^{\mathcal{V}}_{ttt} &= 0.9582\times \left(f_{\rm FCNI}^{\mathcal{V}} f_{\rm FVNI}^{\mathcal{V}}\right)^2\left(\frac{\text{TeV}}{\Lambda}\right)^4 ~ \rm{pb},\nonumber\\
\sigma^{\mathcal{S}}_{ttt} &= 0.3131 \times \left(f_{\rm FCNI}^{\mathcal{S}} f_{\rm FVNI}^{\mathcal{S}}\right)^2\left(\frac{\text{TeV}}{\Lambda}\right)^4 ~ \rm{pb}.
\end{align}
Figure~\ref{fig:xsec} displays the cross section $\sigma^{\mathcal{V(S)}}_{ttt}$ as a function of  $\Lambda$ for $f_{\rm FCNI}^{\mathcal{V(S)}} f_{\rm FVNI}^{\mathcal{V(S)}}=1$ at the 13~TeV LHC.

The best way to measure the triple top-quark events is through the same-sign charged-lepton mode, which demands the same-sign top-quark pairs decay leptonically while the third top-quark decays hadronically, i.e.
\begin{align}
u g \rightarrow t t \bar{t}, \quad t \rightarrow b l^+ \nu , \quad \bar{t} \rightarrow \bar{b} j j; \\
\bar{u} g \rightarrow \bar{t} \bar{t} t, \quad \bar{t} \rightarrow \bar{b} l^- \bar{\nu}, \quad t \rightarrow b j j .
\end{align}
The sample of events of interest to us is characterized by two high-energy same-sign leptons, multiple $b$-jets, light-flavor jets and a large missing transverse momentum ($\met$) arising from the invisible neutrinos in the final state. 
The dominant backgrounds yielding the same collider signature are the process of the $t\bar{t}V$ productions ($V=W/Z$) and the $t\bar{t}$ pair production. The first process ($t\bar{t}V$) is the SM irreducible background while the second ($t\bar{t}$) is a reducible background as it contributes when some particles are mis-tagged. For example, one of the $b$-quarks decays into an isolated charged lepton while one of the two jets from the $W$-boson decay is mis-tagged as a $b$-jet.

\begin{figure}[b]
\includegraphics[scale=0.2]{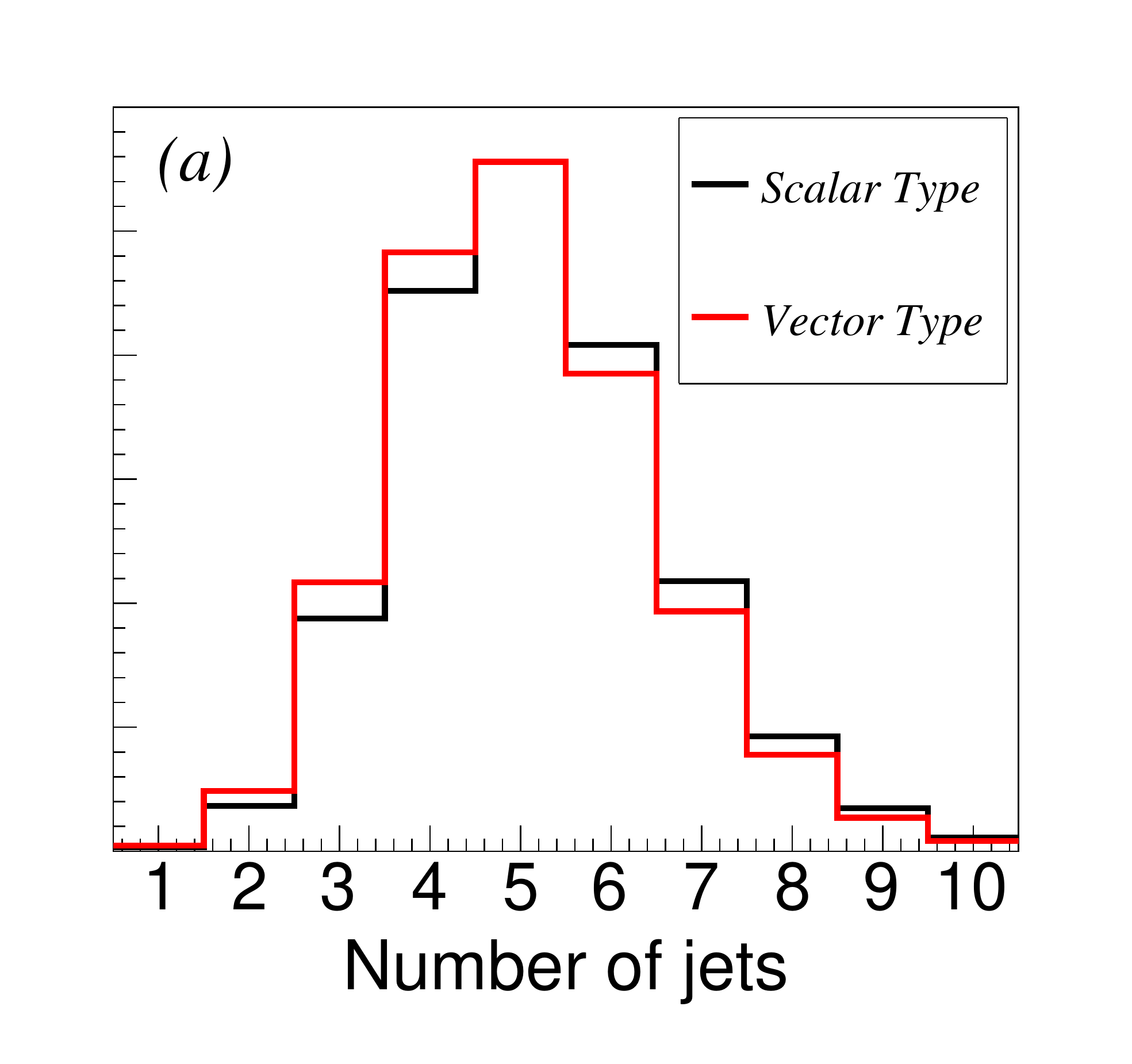}
\includegraphics[scale=0.2]{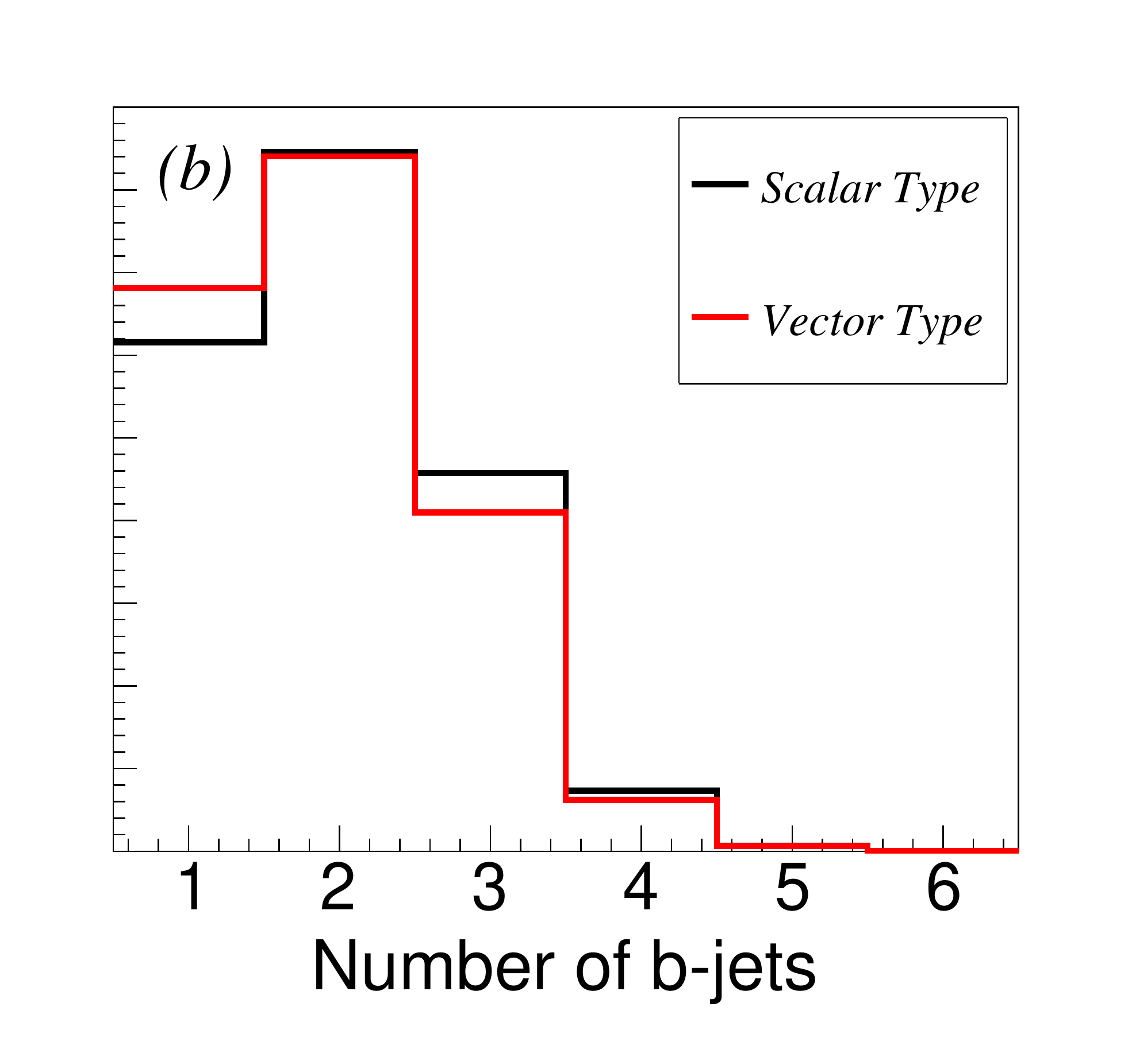}
\includegraphics[scale=0.2]{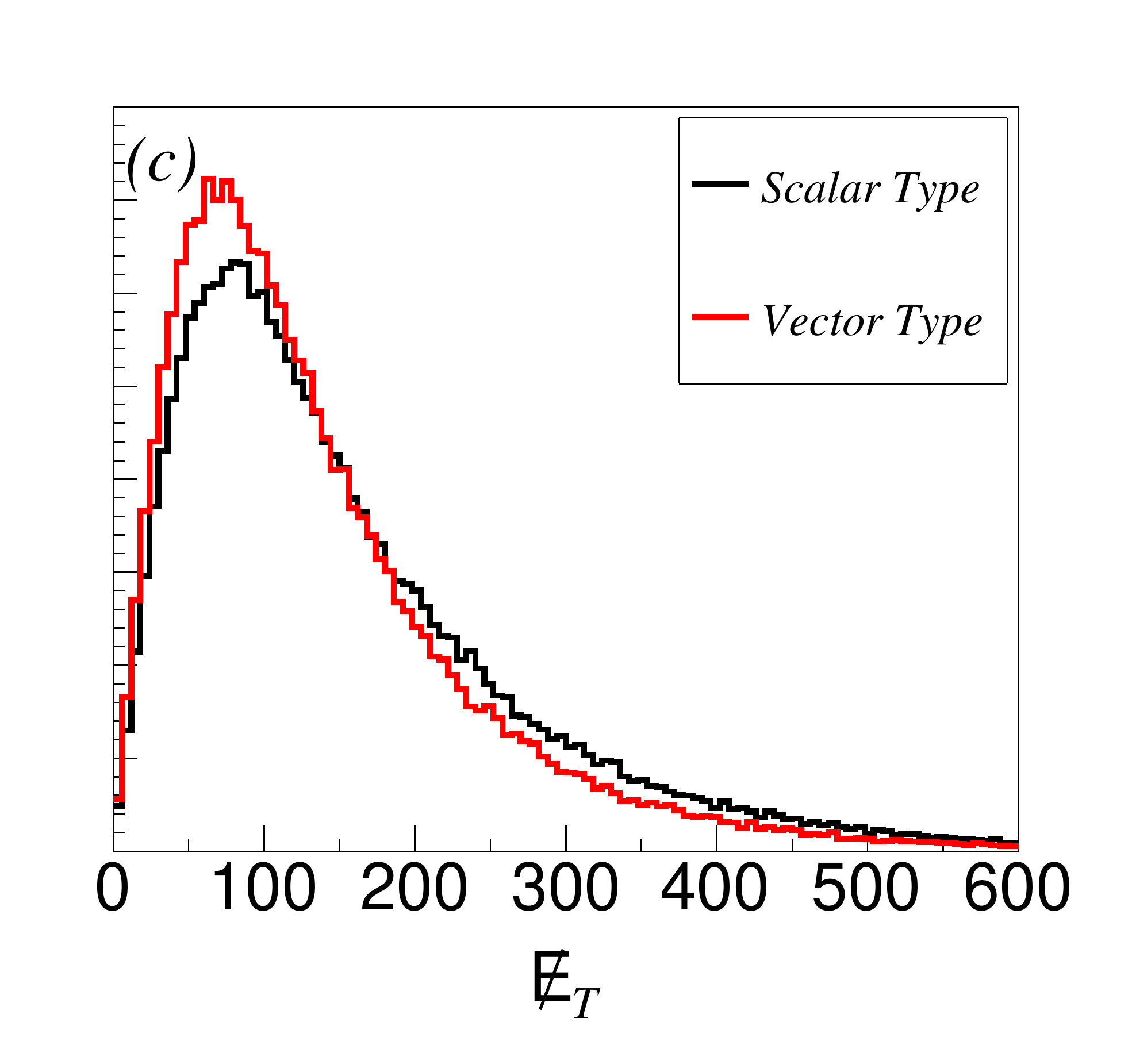}
\includegraphics[scale=0.2]{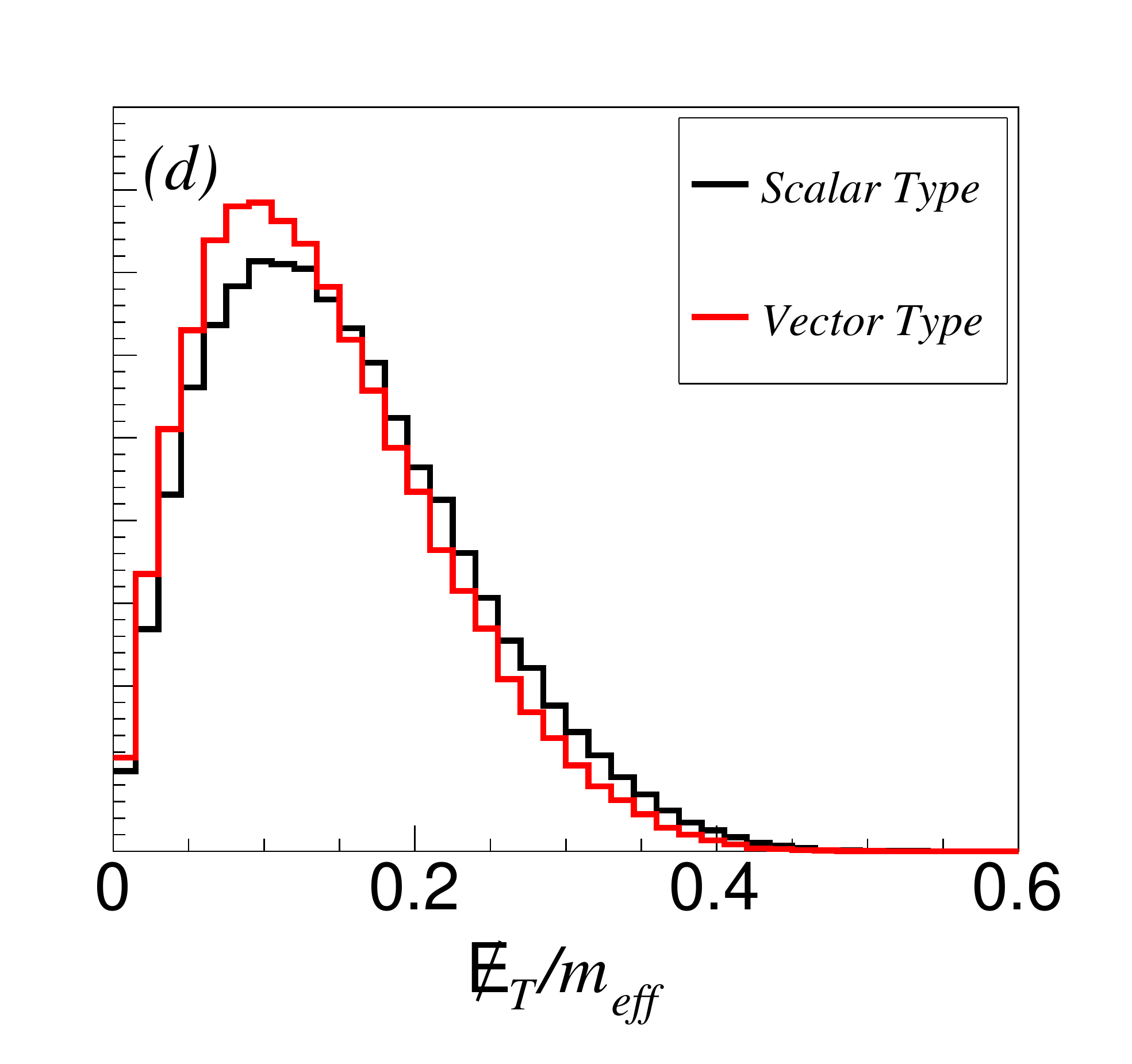}
\caption{\it Normalized distributions of various kinematics variables after demanding at least two same-sign leptons.}
\label{ttt_kin} 
\end{figure}

Both the ATLAS and the CMS collaborations have searched for NP signals with the signature of same-sign leptons and multiple jets~\cite{Chatrchyan:2013fea,Aad:2014pda,Aaboud:2017dmy,Sirunyan:2017uyt}. Based on a data sample corresponding to an integrated luminosity ($\mathcal{L}$) of 36.1 fb$^{-1}$ at the 13~TeV LHC, the ATLAS group reports several signal regions based on the corresponding NP topology~\cite{Aaboud:2017dmy}, 
e.g. an optimal signal region (named as Rpc2L1bH) is defined as follows:
\begin{align}
&N_{\ell^+(\ell^-)}=2, &&N_{b}\geq 1, && N_{\rm jets}\geq 6, \nonumber\\
&\met>250~{\rm GeV}, &&\met/m_{\rm eff} > 0.2~,
\label{eq:cuts}
\end{align}
where $N_{\ell^+(\ell^-)}$ ($N_b$, $N_{\rm jets}$) denotes the number of same-sign leptons ($b$-jets, light-flavor jets), respectively. $m_{\rm eff}$ is defined as the scalar sum of transverse momenta of all the visible particles in the final state and the missing transverse momentum. 

We employ the searching strategy used by the ATLAS collaboration and explore the potential of the LHC on the triple top-quark production. We generate the signal and background events using MadGraph5 \cite{Alwall:2014hca} and then link them with Pythia~\cite{Sjostrand:2006za} and Delphes~\cite{deFavereau:2013fsa} for parton shower, hadronization and detector simulation. Figure~\ref{ttt_kin} displays a few normalized distributions of the signal event after imposing same-sign lepton pair cut: (a) the numbers of $b$-jets; (b)  the numbers of jets; (c) $\met$; (d) the ratio $\met/m_{\rm eff}$. The black and red curves denote the distributions of the scalar operator $\mathcal{O}^{\mathcal{S}}_{uttt}$ and the vector operator $\mathcal{O}^{\mathcal{V}}_{uttt}$, respectively.  Both type of operators yield almost identical distributions. We observe that 0.14\% of the signal events passing the optimal cuts for the vector operator while 0.24\% for the scalar operator. 
The ATLAS collaboration shows that only 9.8 events of the SM background survive the optimal cuts at the 13~TeV LHC with $\mathcal{L}= 36.1~{\rm fb}^{-1}$~\cite{Aaboud:2017dmy}. The numbers of background events ($n_b$) at other integrated luminosities can be obtained by the simple rescaling $n_b(\mathcal{L})=9.8\times (\mathcal{L}/36.1)$.

\begin{figure}
\includegraphics[scale=0.4]{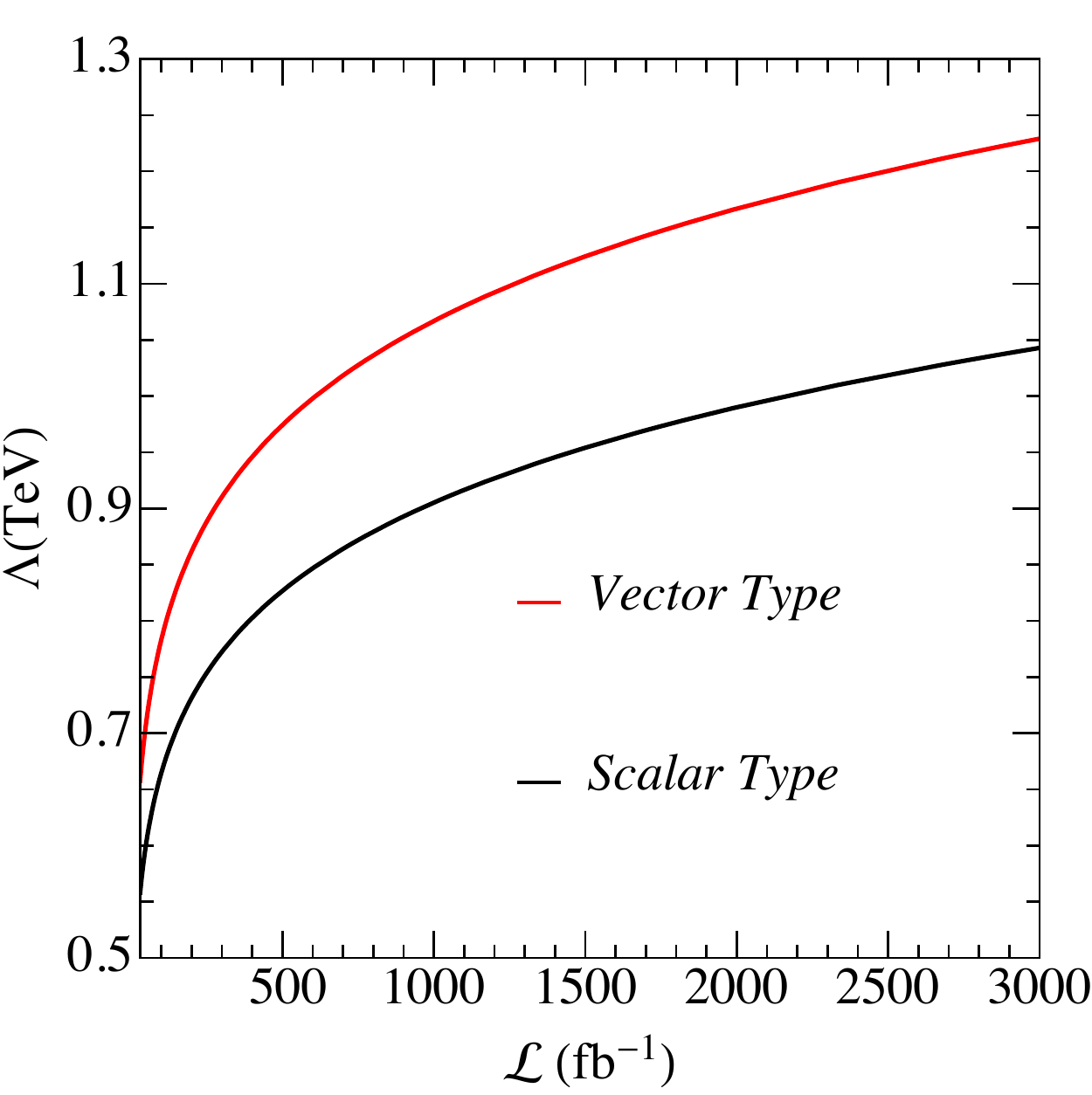}
\caption{\it 95\% C.L. lower bounds on $\Lambda$ as a function of $\mathcal{L}$ for $f_{\rm FCNI}^{\mathcal{S}} f_{\rm FVNI}^{\mathcal{S}} =f_{\rm FCNI}^{\mathcal{V}} f_{\rm FVNI}^{\mathcal{V}}=1$ at the 13~TeV LHC.}
\label{lambda_lumi}
\end{figure}

Equipped with the cut efficiency of the signal and the event number of the SM background, we get the exclusion region of the scale $\Lambda$ and Wilson coefficients at a 2 standard deviations ($\sigma$) statistical significance in terms of
\begin{equation}
\sqrt{-2\left[n_b\log \frac{n_s + n_b}{n_b}-n_s\right]} = 2~,
\label{eq:ex}
\end{equation}
where $n_s$ is the event number of the signal, which is 
\begin{equation}
n_s(\mathcal{L}) = \sigma_{ttt}^{\mathcal{S}(\mathcal{V})}{\rm Br}^2(W^\pm \to \ell^\pm\nu)\times {\rm Br}(W^\mp \to jj) \times \epsilon_{\rm cut}\times \mathcal{L},
\label{eq:ns}
\end{equation}
where $\ell=e/\mu$ and $\epsilon^{\mathcal{S},\mathcal{V}}_{\rm cut}$ denotes how often a signal event would pass the kinematics cuts shown in Eq.~\ref{eq:cuts}, i.e. $\epsilon^{\mathcal{V}}_{\rm cut}=0.14\%$ and  $\epsilon^{\mathcal{S}}_{\rm cut}=0.24\%$. 
We thus obtain 95\% C.L. upper bounds on the Wilson coefficients at the 13~TeV LHC with $\mathcal{L}=300~(3000)~\text{fb}^{-1}$ as follows:
\bea
&& f^{\mathcal{V}}_{\rm FVNI} f^{\mathcal{V}}_{\rm FCNI} \leq 1.21~(0.66)\times \left(\frac{\Lambda}{\text{TeV}}\right)^2, \nn\\
&& f^{\mathcal{S}}_{\rm FVNI} f^{\mathcal{S}}_{\rm FCNI}\leq 1.62~(0.89) \times \left(\frac{\Lambda}{\text{TeV}}\right)^2.
\label{eq:ttt2sigma}
\eea
respectively. Setting $f_{\rm C}^{\mathcal{S}} f_{\rm V}^{\mathcal{S}} =f_{\rm C}^{\mathcal{V}} f_{\rm V}^{\mathcal{V}}=1 $ gives rise to lower bounds on $\Lambda$ as follows:
\beq
 \Lambda^{\mathcal{S}}\geq 773~(1043)~\text{GeV},  \quad \Lambda^{\mathcal{V}}\geq 910~(1229)~\text{GeV}.
\eeq
Figure~\ref{lambda_lumi} displays lower bounds on $\Lambda$ at 95\% C.L. as a function of the integrated luminosity $\mathcal{L}$ for $f_{\rm C}^{\mathcal{S}} f_{\rm V}^{\mathcal{S}} =f_{\rm C}^{\mathcal{V}} f_{\rm V}^{\mathcal{V}}=1 $.

We also vary the Wilson coefficients to obtain the parameter space of discovering an excess in the triple-top production with a $5\sigma$ statistical significance using
\beq
\sqrt{-2\left[(n_b + n_s) \log\frac{n_b}{n_s+n_b}+n_s\right]}=5.
\label{eq:5sigma}
\eeq
We obtain the discovery regions of triple-top productions at the 13~TeV LHC with $\mathcal{L}=300~\text{fb}^{-1}$ ($3000~{\rm fb}^{-1}$) as follows:
\bea
&& f^{\mathcal{V}}_{\rm FVNI} f^{\mathcal{V}}_{\rm FCNI} \geq 1.93~(1.05) \times \left(\frac{\Lambda}{\text{TeV}}\right)^2, \nn\\
&&f^{\mathcal{S}}_{\rm FVNI} f^{\mathcal{S}}_{\rm FCNI}\geq 2.57~(1.14) \times \left(\frac{\Lambda}{\text{TeV}}\right)^2,
\eea
respectively.

~\\
\noindent{\bf 3. $tt/\bar{t}\bar{t}$ and $t\bar{t}t\bar{t}$ productions.}

\begin{figure}
\centering 
\includegraphics[scale=0.21]{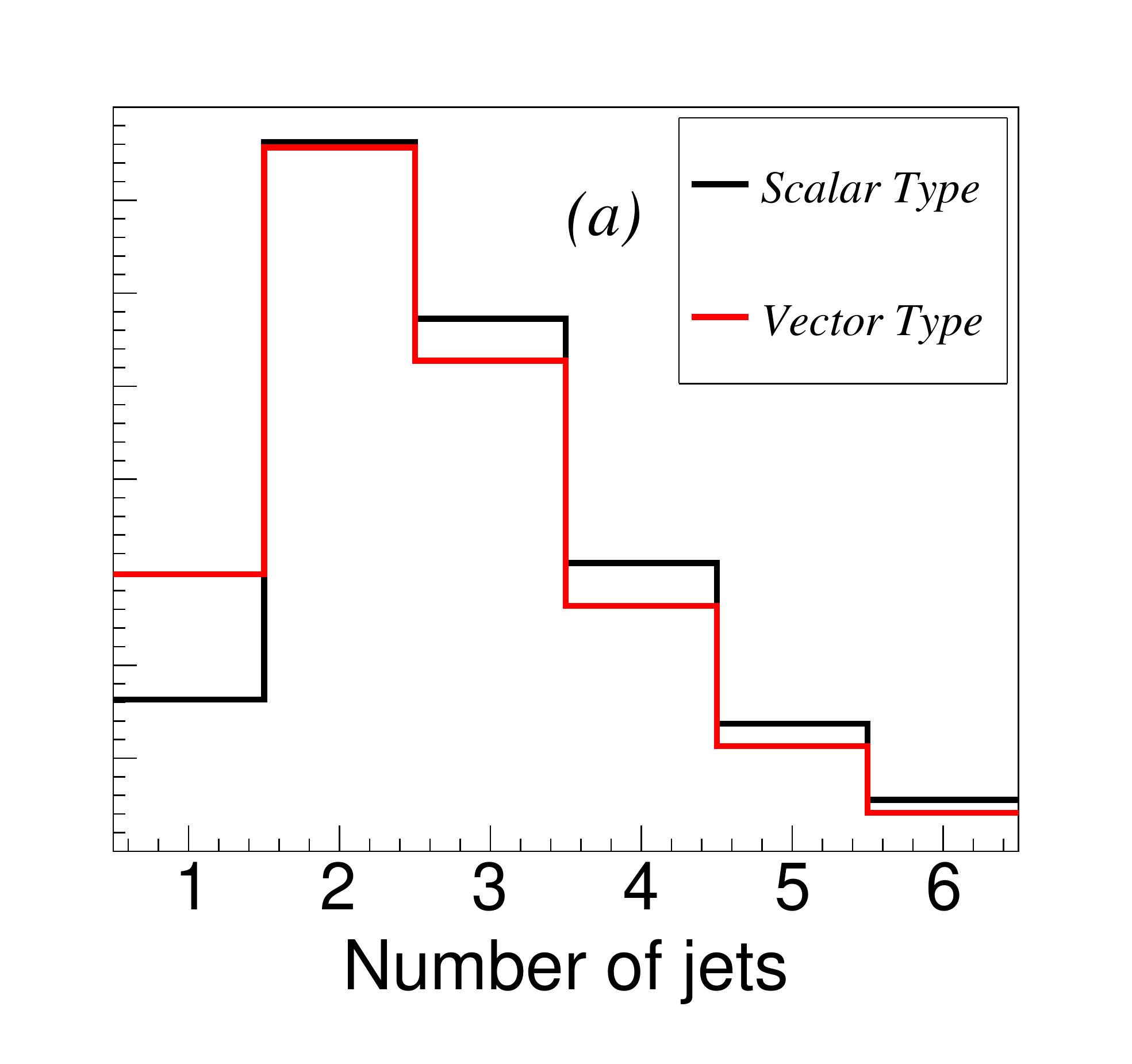}
\includegraphics[scale=0.21]{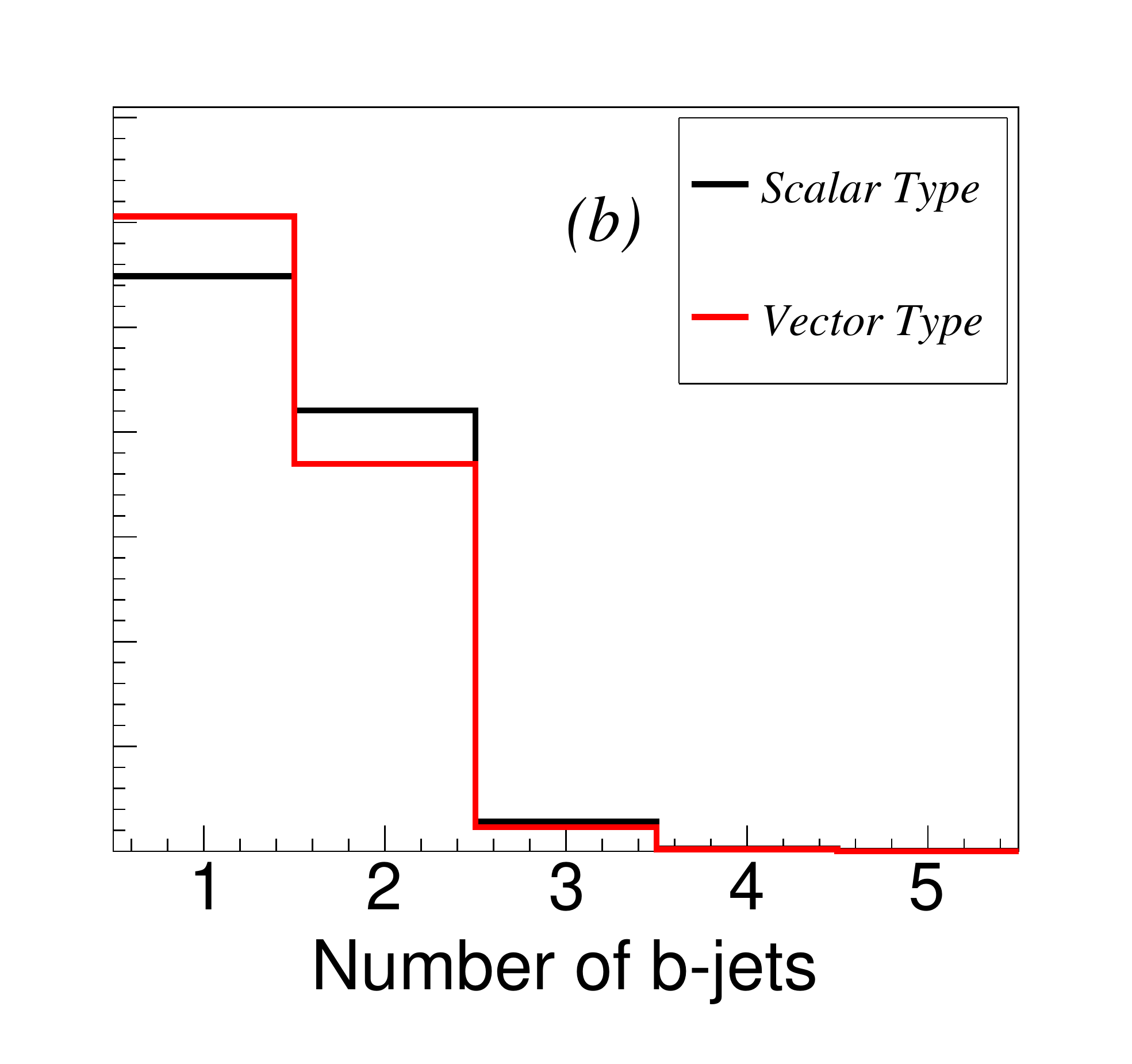}
\caption{\it Numbers of jets and $b$-tagged jets in the $tt/\bar{t}\bar{t}$ channel. }
\label{sst_dis}
\end{figure}

Next we consider the constraints from the same-sign top-quark pair production which involves the FVNI operators $\mathcal{O}^{\mathcal{V},\mathcal{S}}_{uutt} $. 
Similar to the triple-top production, the $tt/\bar{t}\bar{t}$ channel also exhibits a pair of same-sign charged leptons in the final state but with fewer jets and $b$-jets; see Fig.~\ref{sst_dis}. 
We follow the ATLAS collaboration~\cite{Aad:2014pda} to focus on a signal region named as SR1b which is defined as follows:
\begin{align}
&N_{\text{jets}} \ge 3, &&N_{\text{b}-\text{jets}} \ge 1, && \met \ge 150~\text{GeV},\nn \\
& m_T \ge 100~\text{GeV},  &&m_{\text{eff}} \ge 700~\text{GeV}. && 
\label{eq:SR1b}
\end{align}
Here $m_T$ denotes the transverse mass of the leading charged lepton ($\ell_1$) and the missing energy $\met$, defined as $m_T = \sqrt{2p_T^{\ell_1}\met (1 - \cos \Delta \phi)}$ 
where $\Delta\phi$ is the azimuthal angle between the $\ell_1$ lepton and the $\met$. The $m_T$ cut is to remove those backgrounds involving  leptonically decayed $W$-bosons. The $m_{\rm eff}$ is the  the scalar sum of the transverse momenta of all the visible particles and the $\met$.  After all the SR1b cuts there are 4.5 background event at the 13~TeV with $\mathcal{L}=3.2~\text{fb}^{-1}$~\cite{Aad:2014pda}. We find that 0.289\% (0.488\%) of the signal events passing the optimal cuts for the vector (scalar) operator, respectively.

The $tt/\bar{t}\bar{t}$ production cross-section at the 13~TeV LHC can be parameterized as follows:
\bea
&& \sigma^{\mathcal{V}}_{tt+\bar{t}\bar{t}} = 52.28\times (f_{\rm FVNI}^{\mathcal{V}})^4 \left(\frac{\text{TeV}}{\Lambda}\right)^4~\rm{pb},\nn\\
&& \sigma^{\mathcal{S}}_{tt+\bar{t}\bar{t}} = 3.88\times (f_{\rm FVNI}^{\mathcal{S}})^4 \left(\frac{\text{TeV}}{\Lambda}\right)^4~\rm{pb}.
\eea
As no excesses were reported in the $tt/\bar{t}\bar{t}$ channel, we derive the $2\sigma$ bounds on $f_{\rm FVNI}^{\mathcal{V}(\mathcal{S})}$ based on the ATLAS result ($\mathcal{L}=3.2~{\rm fb}^{-1}$)~\cite{Aad:2014pda} as follows:
\beq
f_{\rm FVNI}^{\mathcal{V}}\leq 0.70~\frac{\Lambda}{\text{TeV}},\quad f_{\rm FVNI}^{\mathcal{S}}\leq 1.17~\frac{\Lambda}{\text{TeV}}.
\eeq
The projected $2\sigma$ upper limits on $f_{\rm FVNI}^{\mathcal{V}(\mathcal{S})}$ at the LHC with $\mathcal{L}=300~{\rm fb}^{-1}$ ($3000~{\rm fb}^{-1}$) are 
\beq
f_{\rm FVNI}^{\mathcal{V}}\leq 0.37~(0.28)~\frac{\Lambda}{\text{TeV}},~f_{\rm FVNI}^{\mathcal{S}}\leq 0.62~(0.47)~\frac{\Lambda}{\text{TeV}},
\eeq
respectively, while the $5\sigma$ discovery regions are given as follows:
 \beq
f_{\rm FVNI}^{\mathcal{V}}\geq 0.46~(0.35)~\frac{\Lambda}{\text{TeV}},~f_{\rm FVNI}^{\mathcal{S}}\geq 0.92~(0.69)~\frac{\Lambda}{\text{TeV}}.
\eeq

Last but not the least, we consider the four top-quark production of which the production cross-sections at the 13~TeV LHC are  
\bea
&& \sigma^{\mathcal{V}}_{tt\bar{t}\bar{t}} = 1.166\times (f_{\rm FCNI}^{\mathcal{V}})^4 \left(\frac{\text{TeV}}{\Lambda}\right)^4~\rm{fb},\nn\\
&& \sigma^{\mathcal{S}}_{tt\bar{t}\bar{t}} = 0.804\times (f_{\rm FCNI}^{\mathcal{S}})^4 \left(\frac{\text{TeV}}{\Lambda}\right)^4~\rm{fb}.
\eea
The CMS collaborations have searched the process at the 13 TeV LHC with the integrated luminosity of 35.9 fb$^{-1}$ and obtained the upper limit on the four top-quark production cross section of 41.7 fb at the $2\sigma$ CL.~\cite{Sirunyan:2017roi}. In the SM the cross section of the $t\bar{t}t\bar{t}$ production is 9.2 fb after including the next-to-leading-order QCD corrections~\cite{Bevilacqua:2012em,Alwall:2014hca,Frederix:2017wme}. We conclude that the NP contribution to the four-top production cross-section at the 13~TeV is less than 32.5~fb at the $2\sigma$ C.L.,  which yields
\beq
f_{\rm FCNI}^{\mathcal{V}}\leq 2.29\frac{\Lambda}{\text{TeV}}, \quad f_{\rm FCNI}^{\mathcal{S}}\leq 2.51\frac{\Lambda}{\text{TeV}}. 
\eeq 
If the cross-section is further constrained to be less than twice of the SM value when accumulating an integrated luminosity of $300~{\rm fb}^{-1}$, we then obtain projected bounds as follows:
 \beq
f_{\rm FCNI}^{\mathcal{V}}\leq 1.99\frac{\Lambda}{\text{TeV}}, \quad f_{\rm FCNI}^{\mathcal{S}}\leq 2.19\frac{\Lambda}{\text{TeV}}. 
\label{eq:4tprojection}
\eeq

\begin{figure*}
\includegraphics[scale=0.5]{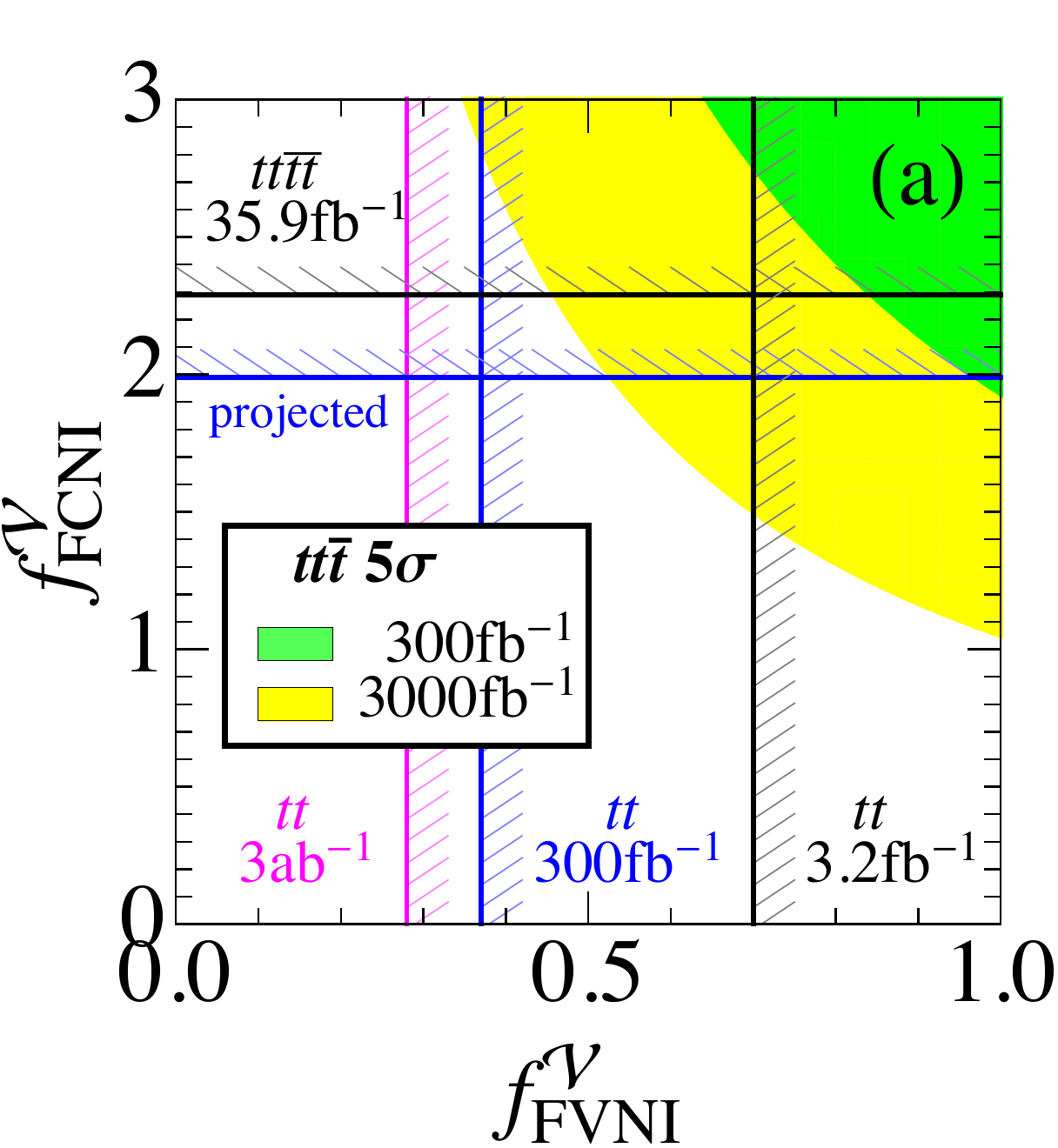}\qquad
\includegraphics[scale=0.48]{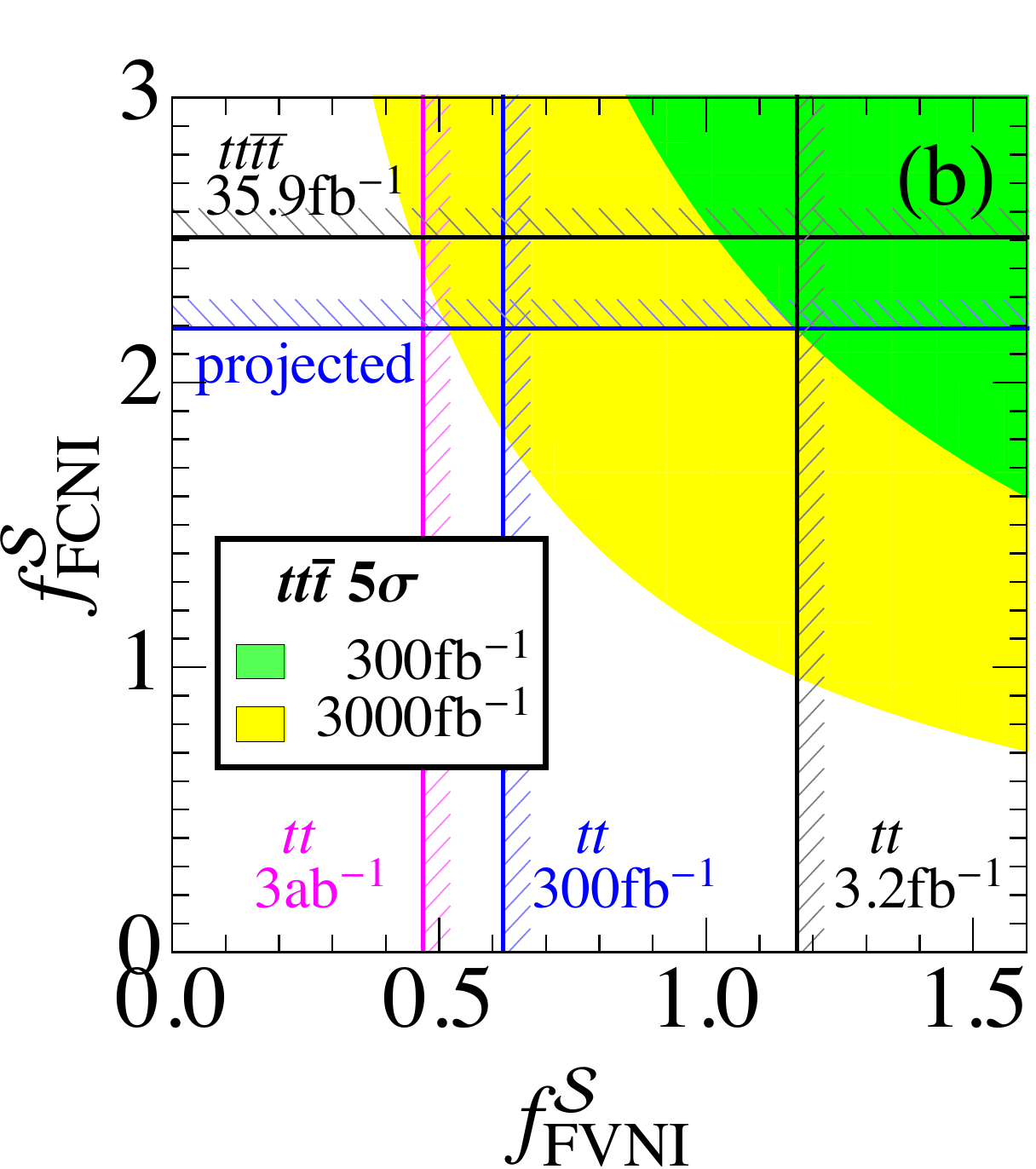}
\caption{\it The potential of the 13~TeV LHC on $f_{\rm FVNI}^{\mathcal{V,S}}$ and $f_{\rm FCNI}^{\mathcal{V,S}}$ when the operators are induced by a heavy vector resonance (a) or by a scalar resonance (b). The green and yellow band denotes the $5\sigma$ discovery region with $\mathcal{L}=300~\text{fb}^{-1}$ and $3000~\text{fb}^{-1}$, respectively. The vertical (horizontal) lines with meshed region denote the exclusion limits from the $tt/\bar{t}\bar{t}$ ($t\bar{t}t\bar{t}$) production, respectively. The black lines represent the current limits while the blue and magenta lines denote projected limits. }
\label{fig:all}
\end{figure*}

\noindent{\bf 4. Combined analysis and summary.}

We combine the triple-top, $tt/\bar{t}\bar{t}$ and $t\bar{t}t\bar{t}$ channels to explore the potential of probing $f_{\rm FVNI}^{\mathcal{V,S}}$ and $f_{\rm FCNI}^{\mathcal{V,S}}$ at the 13~TeV LHC.

For illustration we choose $\Lambda=1~\text{TeV}$ and compare the sensitivities of the three channels in the plane of $f_{\rm FVNI}^{\mathcal{V,S}}$ and $f_{\rm FCNI}^{\mathcal{V,S}}$; see Fig.~\ref{fig:all}(a) for the results of a vector $\mathcal{V}$ while Fig.~\ref{fig:all}(b) for a scalar $\mathcal{S}$. The green (yellow) shaded region denotes the parameter space to reach a discovery of $tt\bar{t}$ production at the $5\sigma$ C.L. at the 13~TeV with an integrated luminosity of $300~\text{fb}^{-1}$ ($3000~\text{fb}^{-1}$), respectively. The vertical lines shows the $2\sigma$ C.L. bounds on $f_{\rm FVNI}^{\mathcal{V,S}}$ derived from the same-sign top-quark pair production, where the black line represent the current bound while the blue and magenta lines denotes the projected bounds. The meshed region on the right-hand side of each vertical line is excluded. The black horizontal line represents the current bound on $f_{\rm FCNI}^{\mathcal{V,S}}$ obtained from the $t\bar{t}t\bar{t}$ production based on the $35.9~\text{fb}^{-1}$ dataset, while the blue horizontal line denotes the projected bounds given in Eq.~\ref{eq:4tprojection}.

We first consider the operators induced by a heavy vector resonance. The $tt/\bar{t}\bar{t}$ production gives rise to very severe bounds on $f_{\rm FVNI}^{\mathcal{V}}$.  Based on the current bounds from $tt/\bar{t}\bar{t}$ and $t\bar{t}t\bar{t}$ productions, the triple-top production cannot be discovered at the 13~TeV LHC with $\mathcal{L}=300~\text{fb}^{-1}$. The green shaded region in Fig.~\ref{fig:all}(a) is completely ruled out the two black lines. Even worse, the discovery region for $\mathcal{L}=3000~\text{fb}^{-1}$ is ruled out by the projected exclusion limits at $\mathcal{L}=300~\text{fb}^{-1}$. See the yellow region and the two blue lines. We arrive at a rather negative conclusion that, if the triple-top production is induced by a vector resonance alone, i.e. all the operators share the same cutoff scale $\Lambda$, then the $tt/\bar{t}\bar{t}$ production is much easier to see or exclude the NP effect than the triple-top production. However, if we observe an excess in the triple-top production but not in the $tt/\bar{t}\bar{t}$ production, then it is possible that there are more than one heavy vector resonances.

On the other hand, the $tt/\bar{t}\bar{t}$ channel imposes mild constraint on $f_{\rm FVNI}^{\mathcal{S}}$. See the black lines in Fig.~\ref{fig:all}(b). As a result, the green (yellow) shaded region on the lower-left side of the two black (blue) lines can be discovered when accumulating an integrated luminosity of $300~\text{fb}^{-1}$ ($3000~\text{fb}^{-1}$), respectively. Finally, if no excesses were observed in the $tt/\bar{t}\bar{t}$ production, then the entire parameter space of discovering the triple-top production induced by a heavy scalar resonance is ruled out; see the magenta line.  

In summary, the triple top-quark production provides complementary information to the on-going new physics searches in the same-sign top-quark pairs and the four top-quark production. We emphasize that the correlations among the three multiple top-quark channels presented in this study will remain the same for different values of $\Lambda$'s.

~\\
\noindent{\bf Acknowledgements.}

The work is supported in part by the National Science Foundation of China under Grant Nos. 11175069, 11275009, 11422545, 11725520, 11675002, 1163500 11775093, 11805013, the Fundamental Research Funds for the Central Universities under Grant No. 2018NTST09 and in part by the U.S. Department of Energy under Contract No. DE-AC02-06CH11357.

\bibliographystyle{apsrev}
\bibliography{reference}

\end{document}